\documentclass[openright, a4paper, 10pt]{article}
\usepackage[utf8]{inputenc}
\usepackage[T1]{fontenc}
\usepackage[USenglish, german]{babel}
\usepackage{datetime}
\usepackage{graphicx}
\usepackage{cite}
\usepackage{notoccite}
\usepackage{setspace}
\usepackage{newclude}
\usepackage[usenames,dvipsnames]{xcolor}
\usepackage{mathtools}
\usepackage{amssymb}
\usepackage{bbm}
\usepackage{mathbbol}
\mathchardef\mhyphen="2D
\DeclareFontFamily{OT1}{pzc}{}
\DeclareFontShape{OT1}{pzc}{m}{it}{<-> s * [1.10] pzcmi7t}{}
\DeclareMathAlphabet{\mathpzc}{OT1}{pzc}{m}{it}
\usepackage{cancel}

\usepackage[noprefix]{nomencl}

\renewcommand{\nomgroup}[1]{\medskip}
\makenomenclature

\usepackage{booktabs}

\definecolor{shadedcolor}{gray}{0.8}
\definecolor{TFFrameColor}{gray}{.8} 
\definecolor{TFTitleColor}{gray}{0} 
\usepackage{framed}

\usepackage{hyperref}
\hypersetup{
    colorlinks,
    citecolor=black,
    filecolor=black,
    linkcolor=black,
    urlcolor=black
}
\usepackage{tabularx}
\usepackage{parskip}

\usepackage{fancyhdr}
\usepackage[toctitles]{titlesec}
\titleformat{\chapter}[hang]{\bfseries\huge}{\thechapter}{2pc}{}
\titlelabel{\thetitle\quad}   

\makeatother
\let\sectionOld\section
\renewcommand\section[2][\empty]{%
	\boldmath\sectionOld[#1]{#2}\unboldmath%
}

\usepackage{enumerate}

\usepackage{siunitx}
\usepackage{units}
\usepackage{upgreek}
\usepackage{tensor}

\usepackage{pgfplots}
  \pgfplotsset{compat=newest, plot coordinates/math parser=false,  
     tick label style={font=\tiny}, legend style={font=\scriptsize}}
  \newlength\figureheight 
  \newlength\figurewidth 
\usepackage[hang,center]{subfigure}
\pgfplotsset{every axis/.append style={
    scaled x ticks=false,
    scaled y ticks=false,
    yticklabel style={/pgf/number format/.cd,fixed,fixed zerofill,precision=1},
    xticklabel style={/pgf/number format/.cd,fixed,fixed zerofill,precision=1}
    }
 }
\pgfplotsset{
	every colorbar/.append style={try min ticks=5, max space between ticks=18pt, at={(1.05,1)} },
	colorbar/width=0.05\figurewidth,
	every axis/.append style={font=\scriptsize}
}

\usepackage{listings}

\definecolor{javaBlue}{RGB}{42,0.0,255}
\definecolor{javaGreen}{RGB}{63,127,95}
\definecolor{javaPurple}{RGB}{127,0,85}

\newcommand{\eq}[1]{\begin{align}#1\end{align}}

\newcommand{\bs}[1]{\boldsymbol{#1}}
\newcommand{\mc}[1]{\mathcal{#1}}
\newcommand{\field}[1]{\mathbbm{#1}}

\newcommand{\total}{\mathrm{d}}

\newcommand{\tdiff}[2]{\dfrac{\mathrm{d}#1}{\mathrm{d}#2}}

\newcommand{\tx}[1]{\text{#1}}
\newcommand{\mrm}[1]{\mathrm{#1}}

\newcommand{\ie}{i.e.\ }
\newcommand{\eg}{e.g.\ }



\usepackage[hang]{footmisc}

\let\originalleft\left
\let\originalright\right
\renewcommand{\left}{\mathopen{}\mathclose\bgroup\originalleft}
\renewcommand{\right}{\aftergroup\egroup\originalright}

\DeclareMathOperator*{\Motimes}{\text{\raisebox{0.25ex}{\scalebox{0.8}{$\bigotimes$}}}}

\DeclareMathOperator*{\Moplus}{\text{\raisebox{0.25ex}{\scalebox{0.8}{$\bigoplus$}}}}

\begin{document}

	\selectlanguage{USenglish}
	
%
%
%
%

	\singlespacing
	\pagestyle{fancy}
	
	\lhead[\MakeUppercase{IBNM Preprint 09/2015}]{\MakeUppercase{IBNM Preprint 09/2015}}
	\rhead[]{}

	\pagenumbering{roman}

	\setcounter{tocdepth}{3}


%
%
%
	
	\parindent=.5cm
	\selectlanguage{USenglish}

	\selectlanguage{USenglish}
	
	\parindent=.5cm
	\pagestyle{plain}
	
	
%
	
	
	\cleardoublepage
	\pagenumbering{arabic}

	\pagestyle{fancy}

\thispagestyle{plain}

\begin{center}
	{\bf IBNM Preprint 09/2015}\\
	\vspace{.5cm}
	{\LARGE A multilevel adaptive sparse grid stochastic collocation approach to the non-smooth forward propagation of uncertainty in discretized problems}\\
	\vspace{.8cm}
	{\large Robert L.\ Gates{\renewcommand{\thefootnote}{\fnsymbol{footnote}}\footnote{Institute of Mechanics and Computational Mechanics (IBNM)\\Gottfried Wilhelm Leibniz Universität Hannover, Appelstr. 9A, D-30167 Hannover}}\footnote[1]{robert.gates@ibnm.uni-hannover.de}  and Maximilian R.\ Bittens{\renewcommand{\thefootnote}{\fnsymbol{footnote}}\footnotemark[1]}\footnote[2]{maximilian.bittens@ibnm.uni-hannover.de}}\\
	\vspace{.2cm}
\end{center}
\vspace{1cm}

\noindent
\begin{center}{\bf Abstract}\medskip\end{center}
This work proposes a scheme for significantly reducing the computational complexity of discretized problems involving the non-smooth forward propagation of uncertainty by combining the adaptive hierarchical sparse grid stochastic collocation method (ALSGC) with a hierarchy of successively finer spatial discretizations (e.g.\ finite elements) of the underlying deterministic problem. To achieve this, we build strongly upon ideas from the Multilevel Monte Carlo method (MLMC), which represents a well-established technique for the reduction of computational complexity in problems affected by both deterministic and stochastic error contributions. The resulting approach is termed the Multilevel Adaptive Sparse Grid Collocation (MLASGC) method. Preliminary results for a low-dimensional, non-smooth parametric ODE problem are promising: the proposed MLASGC method exhibits an error/cost-relation of $\varepsilon \sim t^{-0.95}$ and therefore significantly outperforms the single-level ALSGC ($\varepsilon \sim t^{-0.65}$) and MLMC methods ($\varepsilon \lesssim t^{-0.5}$).\\
\vspace{.5cm}

\section{Introduction}
It is well-known that many problems in engineering are subject to uncertainty in the input parameters, e.g.\ material data, boundary conditions, and geometry. In the present case, we are interested in the forward-propagation of such uncertainty to the quantity of interest, e.g.\ deformation or stress, which is usually obtained from the solution of a partial differential equation. Three classical categories of methods exist for the solution of such stochastic partial differential equations, namely the Stochastic Galerkin (SG)~\cite{Deb2001}, Stochastic Collocation (SC)~\cite{Babuska2007}, and Monte Carlo (MC)~\cite{Babuska2004} approaches. The selection of an appropriate method depends strongly on the number and independence assumptions of the random variables, as well as on the smoothness of the solution in stochastic state space. Regardless of the employed method, stochastic partial differential equations pose a challenge to being solved efficiently when the underlying deterministic problem is computationally costly. It is the purpose of this work to contribute to this field by proposing a method for reducing the computational complexity for a wide range of moderate-dimensional parametric problems encountered in engineering practice.

In the following, we consider a general class of problems from the field of computational mechanics, where solutions to inequality constrained stochastic partial differential equations are sought. Prominent examples include the computation of complex material behavior with uncertain material parameters as well as the contact of deformable bodies with rough surfaces in a generally nonlinear setting. Due to the complimentary condition, such problems are classified as being non-smooth in the random parameter domain~\cite{Bierig2014}. In both exemplary cases, the rough surface as well as the uncertain material parameter field is represented by a stochastic process which, provided that it has bounded second moment, can be approximated by a truncated Karhunen-Loève expansion. This leads to the assumption of a multilinear combination of independent random variables parametrizing the deterministic problem. Due to independence, it is possible to utilize a double-orthogonal polynomial basis of tensor-product structure for the state space, leading to a decoupling of the random dimensions (see e.g.~\cite{Forster2010, Babuska2007, Babuska2004}). This choice of basis renders the SG and SC approaches equivalent, resulting in a non-intrusive method.

Unlike Monte Carlo integration, the convergence of classical SG and SC methods relies on regularity properties of the quantity of interest in the random parameter domain~\cite{Babuska2007, Beck2014}. Unfortunately, the class of problems considered herein clearly violates the required smoothness assumptions~\cite{Bierig2014}, making MC methods an attractive alternative, despite their slow convergence. However, in many cases, areas of reduced regularity are confined to certain regions of state space, suggesting an error-adaptive approach to sparse grid SC methods~\cite{Ma2009, Ma2010, Gunzburger2014} for problems involving a moderately large number of stochastic dimensions. Such adaptive methods utilize local hierarchical basis functions with tensor product structure, naturally providing for an improvement estimate for each adaptively refined collocation point while overcoming the oscillations incurred by the use of global interpolating polynomials.

This work proposes a scheme for significantly reducing the computational complexity of discretized problems involving the non-smooth forward propagation of uncertainty by combining the adaptive hierarchical sparse grid stochastic collocation method~\cite{Ma2009, Ma2010, Gunzburger2014} with a hierarchy of successively finer spatial discretizations (e.g.\ finite elements) of the underlying deterministic problem. To achieve this, we build strongly upon ideas from the Multilevel Monte Carlo method (MLMC)~\cite{Heinrich1998, Heinrich2000, Heinrich2001, Giles2008}, which represents a well-established technique for the reduction of computational complexity in problems affected by both deterministic and stochastic error contributions. The resulting approach is termed the Multilevel Adaptive Sparse Grid Collocation (MLASGC) method. It is remarked that previous works on the topic of multilevel methods in adaptive sparse grid stochastic collocation methods~\cite{Galindo2015} focus on the acceleration of iterative solvers by using a low-fidelity interpolant of the stochastic state space as an initial guess for newly added collocation points. It is emphasized that, while the underlying ideas are very similar, our approach is more classical in that we consider the term ``multilevel'' to apply to a hierarchy of successively finer spatial discretizations, as suggested in~\cite{Wyk2014}.

\section{The multilevel adaptive sparse grid collocation method}
We begin by summarizing the so-called Adaptive Lagrangian Sparse Grid Collocation method (ALSGC)~\cite{Klimke2005, Ma2009, Ma2010} and subsequently extend it to the use of multilevel deterministic discretizations. 

\subsection{Construction of an initial sparse grid}
For the construction of a $d$-dimensional sparse grid on $I_1 \times \ldots \times I_j \times \ldots \times I_d$ by the Smolyak algorithm, introduce the multi-index $\bs{i} = (i_1, \ldots, i_j, \ldots, i_d)$, in which $i_j \in \field{N}^+$. Further, define $\mc{S}_l = \{\bs{i} \;|\; 1-d+|\bs{i}| = l\}$ as the set of multi-indexes belonging to level $l$ of the sparse grid. In order to relate the level $i_j$ of the $j$-th dimension to a certain number of univariate points, we choose a nested rule, \eg the Clenshaw-Curtis rule:
\eq{
	n(i_j) = 
	\begin{cases}
		1 & \text{if} \; i_j = 1\\
		2^{i_j-1}+1 & \text{if} \; i_j > 1  \, 
	\end{cases}
}
and denote $\Theta_{i_j}$ as the resulting set of $n(i_j)$ univariate points on the interval $I_j$. For each level $l$, the sparse grid is constructed by the relation
\eq{
	\bs{\Theta}_l = \bigcup\limits_{\bs{i} \in \mc{S}_l} (\Theta_{i_1} \times \ldots \times \Theta_{i_j} \times \ldots \times \Theta_{i_d}) \, .
}
The adaptive grid discussed in the following section requires an initial set of sparse grid points up to level $L_\tx{init}$ as a starting point for refinement. Hence, we define the initial grid $\bs{\Theta}_\tx{init}$ as the union of the incremental, \ie level-wise, grids $\bs{\Theta}_l$:
\eq{
	\bs{\Theta}_\tx{init} = \bigcup\limits_{l = 1}^{L_\tx{init}} \bs{\Theta}_l \, .
}

\subsection{Tree structure of sparse grid points}
Due to the nested structure of the collocation points, the sparse grid introduced in the previous section admits a $k$-ary tree structure. In particular, every parent collocation point on level $l$ has at most two children per dimension on level $l+1$, \ie $d \le k \le 2d$. In order to make this notion precise, we identify univariate points on level $i_j$ of the $j$-th dimension by integers $m^{i_j}_j \in \{ m \in \field{N}^+ : m \le n(i_j )\}$. Given a univariate parent collocation point on level $i_j$ identified by the index $m^{i_j}_j$, its index on level $i_j + 1$ is given by
\eq{
	m^{i_j+1}_j = 
	\begin{cases}
		2 & \text{if} \; i_j = 1\\
		2 m^{i_j}_j - 1 & \text{if} \; i_j > 1 \, .
	\end{cases}
}
It is emphasized that an existing collocation point identified by the index $m^{i_j}_j$ exists only on level $i_j$, while its index $m^{i_j+1}_j$ on level $i_j+1$ simply corresponds to a non-existent placeholder resulting from the nested structure. This convention allows for the identification of the at most two univariate children on level $i_j+1$ to point $m^{i_j}_j$  as
\eq{
	\bs{c}^{m_j}_{i_j} = \begin{cases}
		\left\{m^{i_j+1}_j + 1\right\} & \text{if} \; m^{i_j+1}_j = 1\vspace{.05cm} \\
		\left\{m^{i_j+1}_j - 1\right\} & \text{if} \; m^{i_j+1}_j = n(i_j+1)\vspace{.05cm}\\
		\left\{m^{i_j+1}_j - 1, m^{i_j+1}_j + 1\right\} & \text{if otherwise}\, .
	\end{cases}
}
For a univariate point identified by the integer $m^{i_j}_j$ as well as the level $i_j$, it is straightforward to recover its coordinate. For instance, in the case of univariate points distributed evenly on the interval $I_j = [-1,1]$, the coordinate is
\eq{
	\hat{\xi}(m^{i_j}_j) = 
	\begin{cases}
	0 & \text{if} \; n(i_j) = 1\\
	-1+\frac{2\left(m^{i_j}_j-1\right)}{n(i_j)-1} & \text{if otherwise} \, .\\
	\end{cases}
} 
Turning back to the multi-dimensional case, collocation points on level $l$ are identified by the multi-index $\bs{m}^{\bs{i}} = (m^{i_1}_1, \ldots, m^{i_j}_j, \ldots, m^{i_d}_d)$, where $\bs{i}\in\mc{S}_l$. We are then able to identify its coordinates 
\eq{
\hat{\bs{\xi}}\left(\bs{m}^{\bs{i}}\right) = \left\{\hat{\xi}\left(m^{i_1}_1\right)\right\} \times \cdots \times \left\{\hat{\xi}\left(m^{i_j}_j\right)\right\} \times \cdots \times \left\{\hat{\xi}\left(m^{i_d}_d\right)\right\}
}
as well as the indices of its children 
\eq{
\bs{c}
^{\bs{m}}_{\bs{i}} = \left\{ \left\{m^{i_1}_1\right\} \times \ldots \times \left\{m^{i_{j-1}}_{j-1}\right\} \times \bs{c}^{m_j}_{i_j} \times \left\{m^{i_{j+1}}_{j+1}\right\} \times \ldots \times \left\{m^{i_{d}}_d\right\} \;|\; j = 1, \ldots, d \right\} \, .
}
Finally, we denote 
\eq{
	\mc{C}_l =
	\begin{cases}
		\left\{ \left\{ \{(1)^d\} \right\} \right\} & \text{if } l = 1\\
		\left\{\bs{c}^{\bs{m}}_{\bs{i}} \;|\; \bs{m}^{\bs{i}} \in  \mc{C}_{l-1} \right\} & \text{if } l > 1
	\end{cases}   
}
as the set of all collocation point indices on level $l$. For an example of the hierarchical construction in two dimensions, see tab.~\ref{tab:tab1}.

\subsection{Local hierarchical Lagrange interpolation}
As already mentioned, the ALSGC approach uses local hierarchical basis functions to overcome the drawback of oscillations, well-known from interpolations using a global Lagrange polynomial basis. We begin with the construction of univariate basis functions, which will then be extended to the multi-dimensional case.

For identifying the support of the hierarchical basis, every point index $m^{i_j}_j$ of an existing collocation point on level $i_{j} > 1$ possesses at most two neighbor point indices
\eq{
	\bs{n}^{m_j}_{i_j} = \begin{cases}
		\left\{m^{i_j}_j + 1\right\} & \text{if} \; m^{i_j}_j = 1 \vspace{.05cm}\\
		\left\{m^{i_j}_j - 1\right\} & \text{if} \; m^{i_j}_j = n(i_j)\vspace{.05cm}\\
		\left\{m^{i_j}_j - 1, m^{i_j}_j + 1\right\} & \text{if otherwise}\, .
	\end{cases}
	\label{eq:nneigh}
}
per dimension $j$. It is remarked that these neighbor point indices need not correspond to existing points. In case of non-existence, they correspond to placeholders in the nested structure. We emphasize that the linear basis function $a^{m_j}_{i_j}$ with significant points 
\eq{
	\bs{p}_{i_j}^{m_j} = \left\{ \hat{\xi}(n) \; | \; n \in \bs{n}^{m_j}_{i_j} \right\} \cup \left\{ \hat{\xi}\left(m^{i_j}_j\right) \right\}
}
has bounded support $\mrm{supp}\left(a^{m_j}_{i_j}\right) = \left[\,\inf \bs{p}_{i_j}^{m_j},\, \sup \bs{p}_{i_j}^{m_j}\,\right]$ and fulfills the following conditions:
\eq{
	a^{m_j}_{i_j}(\xi) = \begin{cases}
		1 & \text{if} \; \xi = \hat{\xi}(m_{i_j}) \\
		0 & \text{if}\,  \xi \not\in \mrm{supp}(a^m_{i_j}) \, .
	\end{cases}
}	
The choice of linear basis functions is made for reasons of simplicity. Other possible candidates include multi-resolution wavelet basis functions~\cite{Gunzburger2014} and higher-order Lagrange polynomials~\cite{Zhang2013}. In all cases, the basis function for level $i_{j}=1$ is defined as 
\eq{
a^1_{1} = 1 \, .
}
Returning to the multidimensional case, the $d$-dimensional basis function can be constructed using tensor products 
\eq{
\bs{a}^{\bs{m}}_{\bs{i}} = a^{m_1}_{i_{1}} \otimes \cdots \otimes a^{m_j}_{i_{j}} \otimes \cdots \otimes a^{m_d}_{i_{d}} = \Motimes\limits_{j=1}^{d} a^{m_j}_{i_{j}}
}
of univariate basis functions. The set of all $d$-dimensional basis functions $\bs{a}^{\bs{m}}_{\bs{i}}$ on level $l$ is denoted by $\mc{A}_{l} = \left\{ \bs{a}^{\bs{m}}_{\bs{i}} \; | \; \bs{m}^{\bs{i}} \in \mc{C}_l \right\}$. For an example of the hierarchical basis in two dimensions, see table~\ref{tab:basis}.

The use of hierarchical basis functions subdivides the sparse grid interpolation space
\eq{
V_{\Gamma} = \Moplus\limits_{l=1}^L W_{l},
}
into an orthogonal sum of hierarchical difference spaces
\eq{
W_{l} = \text{span}\left\{ \bs{a}^{\bs{m}}_{\bs{i}}  : \bs{a}^{\bs{m}}_{\bs{i}} \in \mc{A}_{l} \right\}\, .
}
The interpolant of the hierarchical difference space for a function $f : I^d \rightarrow \field{R}$ is defined as
\begin{align}
\mc{I}_{l}(f)(\bs{\xi}) &= \begin{cases}
\sum\limits_{\bs{m}^{\bs{i}} \in \, \mc{C}_{l}} \bs{a}^{\bs{m}}_{\bs{i}}(\bs{\xi}) \cdot f\left(\hat{\bs{\xi}}\left(\bs{m}^{\bs{i}}\right) \right) & \text{if } l = 1 \\
\sum\limits_{\bs{m}^{\bs{i}} \in \, \mc{C}_{l}} \bs{a}^{\bs{m}}_{\bs{i}}(\bs{\xi}) \cdot \left[ f\left(\hat{\bs{\xi}}\left(\bs{m}^{\bs{i}}\right) \right) - \mc{I}_{l-1}(f)\left(\hat{\bs{\xi}}\left(\bs{m}^{\bs{i}}\right)\right)  \right] & \text{if } l > 1
\end{cases} \\
&= \sum\limits_{\bs{m}^{\bs{i}} \in \, \mc{C}_{l}} {\bs{a}^{\bs{m}}_{\bs{i}}}(\bs{\xi}) \cdot w^{\bs{m}}_{\bs{i}} \, ,
\end{align}
where $w^{\bs{m}}_{\bs{i}}$ denotes the hierarchical surplus belonging to the basis function $\bs{a}^{\bs{m}}_{\bs{i}}$ on level $l$. The hierarchical surplus on level $l=1$ is simply the function value at the coordinates of the collocation point located at $\hat{\bs{\xi}} \left(\bs{m}^{\bs{i}} = (m^1_{1},\ldots,m^1_{j},\ldots,m^1_{d})=(1,\ldots,1,\ldots,1)\right)$. For levels $l>1$, the hierarchical surplus is constructed using the difference of the function value at the coordinates of a collocation point $\hat{\bs{\xi}}\left(\bs{m}^{\bs{i}}\right)$ where $\bs{i} \in \mc{S}_l$ and the value of the interpolation $\mc{I}_{l-1}$ of level $l-1$ at the same coordinates. The complete interpolant is then recovered by the sum over all hierarchical difference interpolants:
\eq{
	\mc{I}^\mrm{SL}(f)(\bs{\xi}) = \sum\limits_{l=1}^L \mc{I}_l(f)(\bs{\xi}) \, .
}

\subsection{Adaptive refinement}
The hierarchical surplus provides for an improvement prediction of the local interpolation when transitioning from level $l$ to $l+1$. It therefore comes naturally to use the hierarchical surplus as the criterion for the adaptive refinement. Hence, for a given tolerance $\epsilon_\mrm{ref} \in \field{R}^{+}$, children $\bs{c}^{\bs{m}}_{\bs{i}}$ of a collocation point on level $l$ identified by the multi-index $\bs{m}^{\bs{i}}$ are created if $|w^{\bs{m}}_{\bs{i}}| > \epsilon_\mrm{ref}$. This allows for the identification of the set of adaptively refined collocation points on level $l+1$ as 
\eq{
	\mc{C}_{l+1} =
		\left\{\bs{c}^{\bs{m}}_{\bs{i}} \;|\; \bs{m}^{\bs{i}} \in \mc{C}_{l} \wedge |w^{\bs{m}}_{\bs{i}}| > \epsilon_\mrm{ref} \right\} \, .
}
It is remarked, that for a proper adaptive refinement the initial level $L_\mrm{init}$ of the sparse grid $\bs{\Theta}_\tx{init}$ must be chosen sufficiently high. Otherwise, the possibility increases that the refinement stops prematurely when a function value at a refined point coincidentally equals the value of the previous level interpolant.
\subsection{Multilevel splitting}
In the most general case, it is assumed that the quantity of interest $u(\cdot, \bs{\xi})$ of the underlying deterministic problem can be approximated to a certain precision level $r$ by an arbitrary scheme, \eg an ordinary differential equation integrated using a difference scheme with time step $\Delta t_r$, a partial differential equation solved via finite elements with characteristic mesh-width $h_r$, or simply a $16\cdot2^r$-bit floating point precision evaluation of a function. In practice, we may be interested in computing the underlying deterministic problem to a precision level $R$, hence it is straightforward to acknowledge that the telescopic sum
\eq{
	u_R = u_{1} + \sum\limits_{r=2}^R u_r - u_{r-1}
}
realizes this requirement. It is remarked that computations to precision $r$ are usually significantly more costly than computations to lesser precision $r-1$.

Due to the hierarchy of successively finer discretizations, we assume that the variance of the level correction $\mathbb{V}[u_r - u_{r-1}] \rightarrow 0$ as $r \rightarrow \infty$. In analogy to MLMC methods, it appears sensible that the number collocation points required for achieving a given interpolation error tolerance is related to the variance of the interpolated function. Hence, the variance decay of the level correction $u_r - u_{r-1}$ is exploited in order to reduce the overall cost of the computation. In particular, we estimate $u_1$ as well as the subsequent corrections $u_r - u_{r-1}$ independently using the ALSGC method. We then recover the response surface approximating $u_R(\bs{\xi})$ by
\eq{
	\mc{I}^\mrm{ML}[u_R](\bs{\xi}) := \mc{I}^\mrm{SL}[u_1](\bs{\xi}) + \sum\limits_{r=2}^R \mc{I}^\mrm{SL}[u_r-u_{r-1}](\bs{\xi}) \, .
}
Subsequent integration of the response surface $\mc{I}^\mrm{ML}[u_R](\bs{\xi})$ for purposes of stochastic moment estimation then follows in a straightforward manner.


\section{Numerical results}

\begin{figure}
	\begin{center}
		\includegraphics{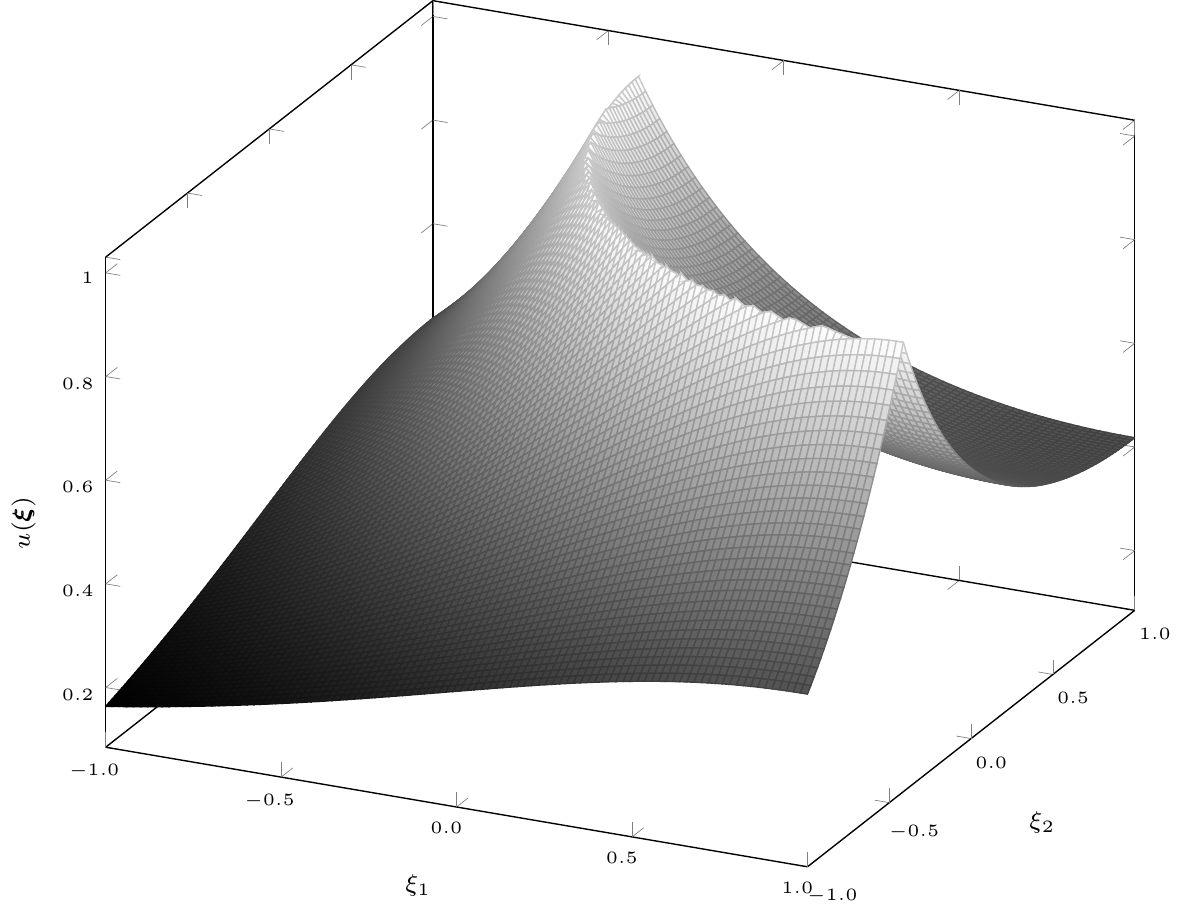}
	\end{center}
	\caption{Exact solution $u(t=1, \bs{\xi})$ of ODE ($\bs{\xi} \in [-1,1]^2$)}
	\label{fig:exactODE}
\end{figure}

\begin{figure}
	\begin{center}
		\includegraphics{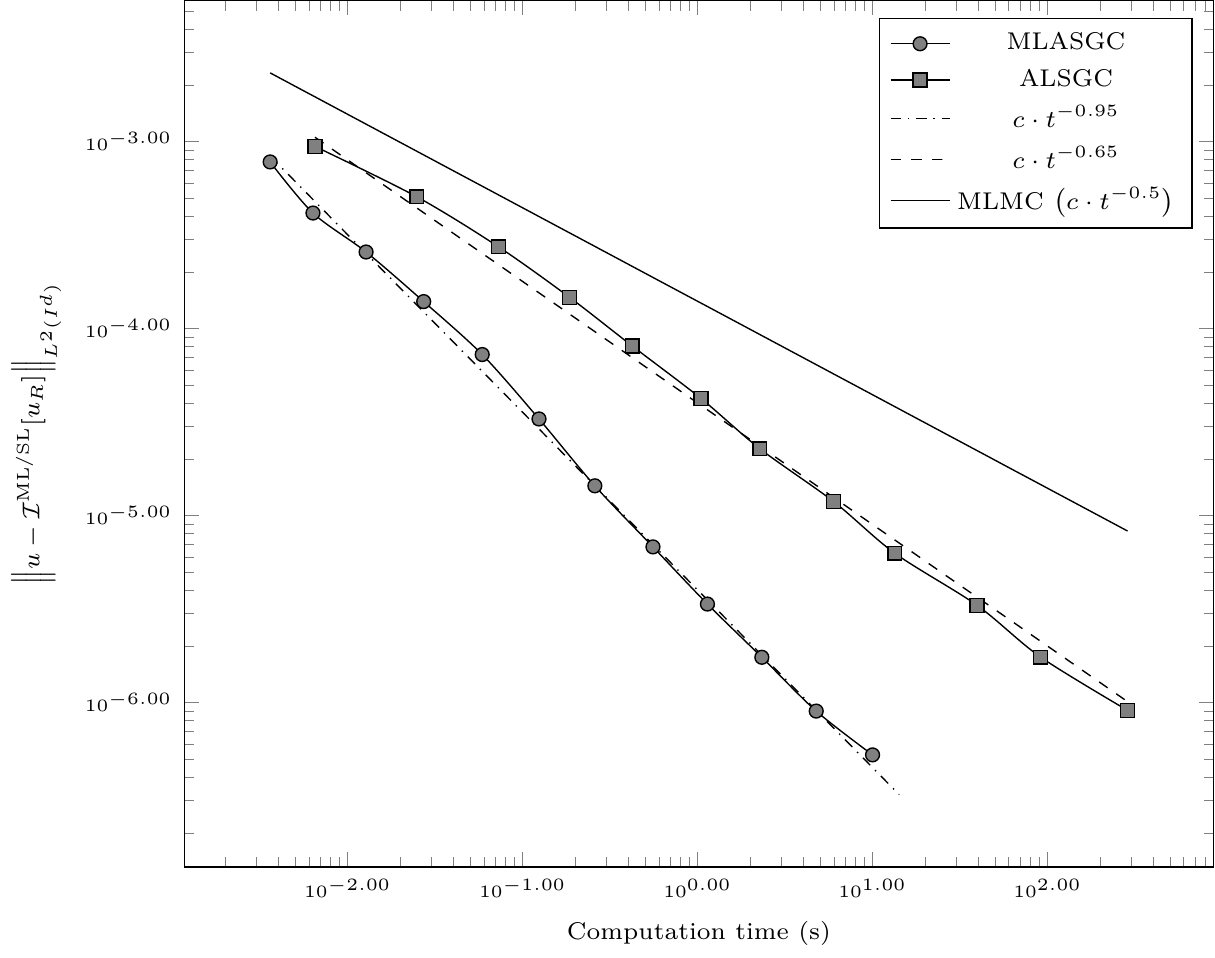}
	\end{center}
	\caption{Error vs.\ cost of the MLASGC, ALSGC, and MLMC methods}
	\label{fig:err}
\end{figure}

\begin{figure}
	\begin{center}
		\includegraphics{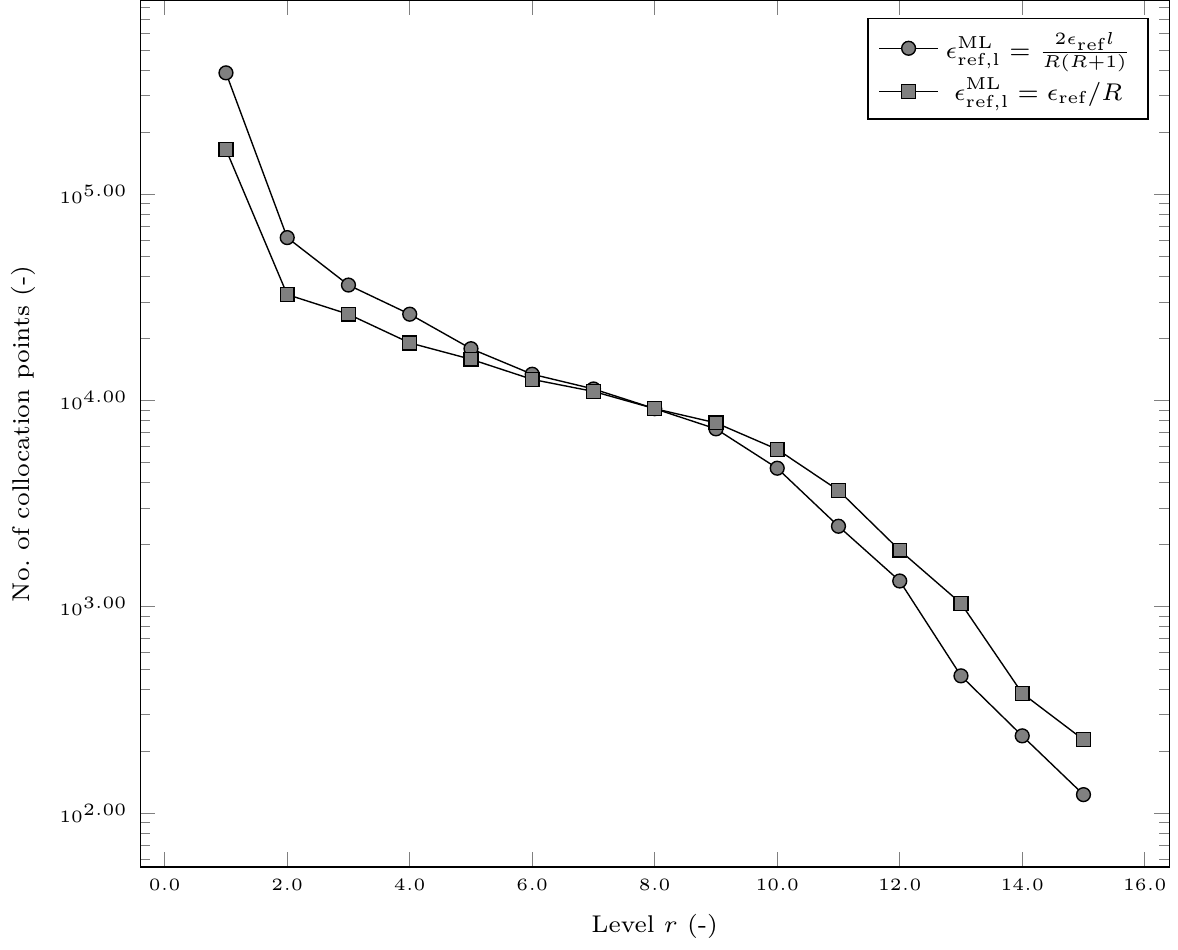}
	\end{center}
	\caption{Number of collocation points needed per discretization level $r$ of the MLASGC method (R=15, $\epsilon_\mrm{ref} = 30^{-1}\cdot2^{-15}$)}
	\label{fig:levelpts}
\end{figure}

For a preliminary numerical investigation, we choose a parametric ($\bs{\xi} \in [-1,1]^2$) first-order linear ordinary differential equation 
\eq{
	\tdiff{u}{t} + \left(|2-(\xi_1-1)^2-(\xi_2-1)^2|+\delta \right) u = 1
}
with initial condition $u(t = 0, \bs{\xi}) = 0$ as well as a regularization parameter $\delta = 10^{-1}$. For reasons of simplicity, we will only be interested in the final value $u(t = 1, \bs{\xi})$. The problem admits an exact solution
\eq{
	u(t, \bs{\xi}) = \frac{1-\exp\left(-t\left(|2-(\xi_1-1)^2-(\xi_2-1)^2|+\delta \right)\right)}{|2-(\xi_1-1)^2-(\xi_2-1)^2|+\delta} \, ,
}
which is shown in fig.~\ref{fig:exactODE} and used as a reference for the numerical solution $u_r(t = 1, \bs{\xi})$, obtained using a forward Euler integration scheme with time-step $\Delta t_r = 30^{-1}\cdot 2^{-r}$. 

We define the total error, including interpolation as well as discretization contributions, in the $L^2$-norm as
\eq{
	\left\| u - \mc{I}^\mrm{ML/SL}[u_R] \right\|_{L^2([-1,1]^2)} = \left(\,\int\limits_{[-1,1]^2} \left|  u(\bs{\xi}) - \mc{I}^\mrm{ML/SL}[u_R](\bs{\xi}) \right|^2 \total \bs{\xi}\right)^{\frac{1}{2}} \, .
}
Figure~\ref{fig:err} shows this error, computed by Monte Carlo integration using $10^5$ points, achieved by the MLASGC and ALSGC methods over computation time. The data points correspond to discretization levels $R=4,\ldots,15$. The adaptive refinement tolerance $\epsilon_\mrm{ref} = \Delta t_R$ for the single-level ALSGC method is chosen equal to the global truncation error $\epsilon_\mrm{discr} \sim \Delta t_R$ of the forward Euler method, aiming to balance the contributions of discretization and interpolation error. For the MLASGC method we distribute the desired overall refinement tolerance $\epsilon_\mrm{ref} = \Delta t_R$ across the $R$ multilevel interpolants in a linear manner such that
\eq{
	\epsilon_\mrm{ref} = \sum\limits_{r=1}^R\epsilon^\mrm{ML}_\mrm{ref,r} = \sum\limits_{r=1}^R \frac{2r\epsilon_\mrm{ref} }{R(R+1)} \, .
}
The linear increase of the refinement tolerance across levels $r$ aims to enforce a high refinement tolerance on coarse discretizations and a lesser refinement tolerance on fine discretizations where additional collocation points would be costly. This choice performed better than the uniform tolerance distribution $\epsilon^\mrm{ML}_\mrm{ref,r} = \epsilon_\mrm{ref}/R$. The idea behind this rather heuristic attempt to balancing cost is confirmed by the number of points required per level $r$ of the multilevel interpolant for achieving an overall refinement error in the order of $\Delta t_R$ (see fig.~\ref{fig:levelpts}). It is however emphasized that even a uniform distribution of the refinement tolerance leads to a significant decay of the required number of collocation points across level corrections, such that the linear distribution should only be considered a slight correction.

\section{Conclusions}
The preliminary results for the low-dimensional, non-smooth parametric ODE problem considered herein are promising: the proposed MLASGC method exhibits an error/cost-relation of $\varepsilon \sim t^{-0.95}$ and therefore significantly outperforms the single-level ALSGC ($\varepsilon \sim t^{-0.65}$) and MLMC methods ($\varepsilon \lesssim t^{-0.5}$~\cite{Cliffe2011, Bierig2014}). Due to a lack of mathematical analysis of the new MLASGC method, no special cost/error-balancing is performed, leaving room for further optimization. It remains to be investigated if the techniques presented in~\cite{Wyk2014, Teckentrup2014} in terms of non-adaptive multilevel collocation methods are similarly applicable to the MLASGC method. It is also emphasized that the new method is not limited to the use of the ALSGC method for the interpolation of the level correction. In fact, a further performance increase would be obtainable by the use of a multi-resolution hierarchical wavelet basis in state space which introduces a true local error estimate due to the fulfillment of the Riesz property (see e.g.~\cite{Gunzburger2014}). Finally, we remark that the non-intrusive nature that the MLASGC method shares with other variants of collocation methods lends itself excellently to parallelization and implementation into existing deterministic frameworks.

\newpage
\section{Appendix}

\subsection{Example: Grid and basis function construction in 2D}
\begin{table}[ht]
	\newcolumntype{L}{>{\centering\arraybackslash} m{.01\textwidth} }
	\newcolumntype{C}{>{\centering\arraybackslash} m{.6\textwidth} }
	\newcolumntype{R}{>{\centering\arraybackslash} m{.25\textwidth} }
\centering
\begin{tabular}{ L | C | R }

$l$ & index-sets $ \mc{S}_l$ and children-sets $\mc{C}_{l}$  & $\bs{m}^{\bs{i}}$  \\
\hline
$1$ & $\begin{array} {lcl} \mc{S}_1 & = & \{\bs{i} \, | \, i_{1}+i_{2}=2 \} \\ & = & \{(1,1)\} \\ \mc{C}_1 & = & \left\{\left\{\left\{(1,1)\right\}\right\}\right\}   \end{array}$  &  \vspace{.1cm}\includegraphics[scale=.8]{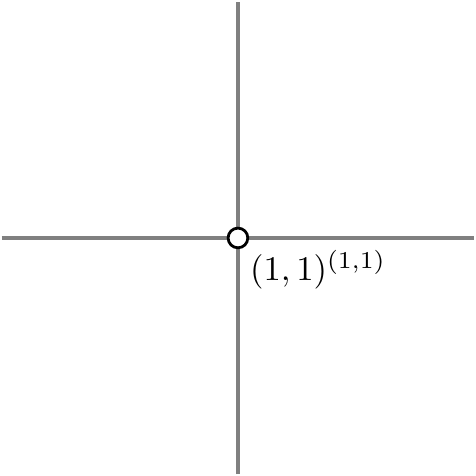}\\
\hline
$2$ & $\begin{array} {lcl} \mc{S}_2  & = & \{\bs{i} \, | \, i_{1}+i_{2}=3 \} \\ & = & \{(2,1),(1,2)\}  \\ \mc{C}_{2}  & = & \left\{ \bs{c}^{\bs{m} = (1,1)}_{\bs{i} = (1,1)} = \{\{1,3\} \times \{1\}, \{1\} \times \{1,3\} \} \right\} \\ & = & \left\{\left\{ \{(1,1),(3,1)\}^{\bs{i}=(2,1)},\{(1,1),(1,3)\}^{\bs{i}=(1,2)} \right\}\right\} \end{array}$  &  \vspace{.05cm}\includegraphics[scale=.8]{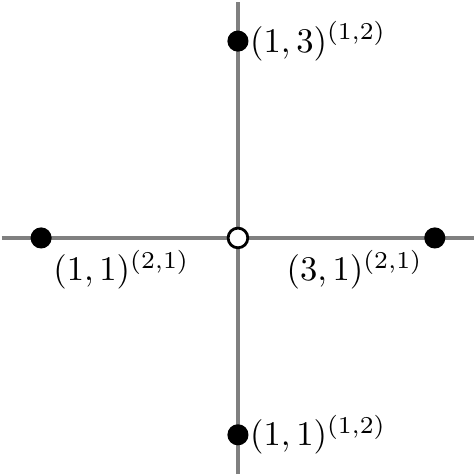}\\
\hline
$3$ & $\begin{array} {lcl} \mc{S}_3  & = & \{\bs{i} \, | \, i_{1}+i_{2}=4 \} \\ & = & \{(3,1), (2,2), (1,3)\}  \\ \mc{C}_{3} & = & \left\{ \bs{c}^{(1,1)}_{(2,1)}, \bs{c}^{(3,1)}_{(2,1)}, \bs{c}^{(1,1)}_{(1,2)}, \bs{c}^{(1,3)}_{(1,2)}  \right\} \\  & = & \big\{  \{ \{2\}\times\{1\}, \{ 1\} \times \{1,3\} \} , \\ && \{ \{4\}\times\{1\}, \{ 3\} \times \{1,3\} \},\\&& \{ \{1,3\}\times\{1\}, \{ 1\} \times \{2\} \},\\&& \{ \{1,3\}\times\{3\}, \{ 1\} \times \{4\} \}\big\}\\ &=&  \big\{  \{ \{(2,1)\}^{(3,1)}, \{ (1,1), (1,3) \}^{(2,2)} \} , \\ && \{ \{(4,1)\}^{(3,1)}, \{ (3,1), (3,3)\}^{(2,2)} \},\\&& \{ \{(1,1),(3,1)\}^{(2,2)}, \{ (1,2)\}^{(1,3)} \},\\&& \{ \{(1,3),(3,3)\}^{(2,2)}, \{ (1,4)\}^{(1,3)} \}\big\} \end{array}$  &  \vspace{.05cm}\includegraphics[scale=.8]{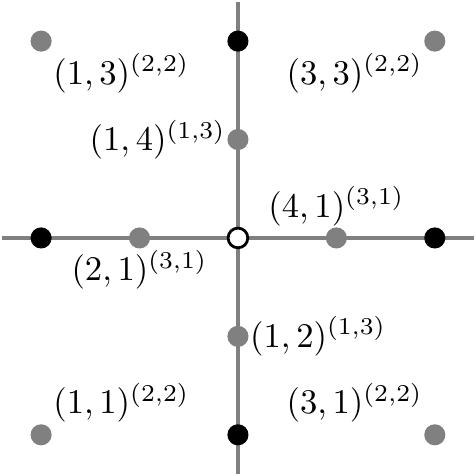}
\end{tabular}
\vspace{.2cm}
\caption{Example: Construction of the sparse grid level $\bs{\Theta}_l$, index-sets $\mc{S}_l$ and children-sets $\mc{C}_l$ in 2D}
\label{tab:tab1}
\end{table}

\begin{table}[ht]
	\newcolumntype{L}{>{\centering\arraybackslash} m{.01\textwidth} }
	\newcolumntype{C}{>{\centering\arraybackslash} m{.9\textwidth} }

	\centering
	\begin{tabular}{ L | C}
		$l$ &basis functions $\bs{a}^{\bs{m}}_{\bs{i}}$ \\
		\hline
		
		1 & \vspace*{.2cm}\includegraphics[scale=.22]{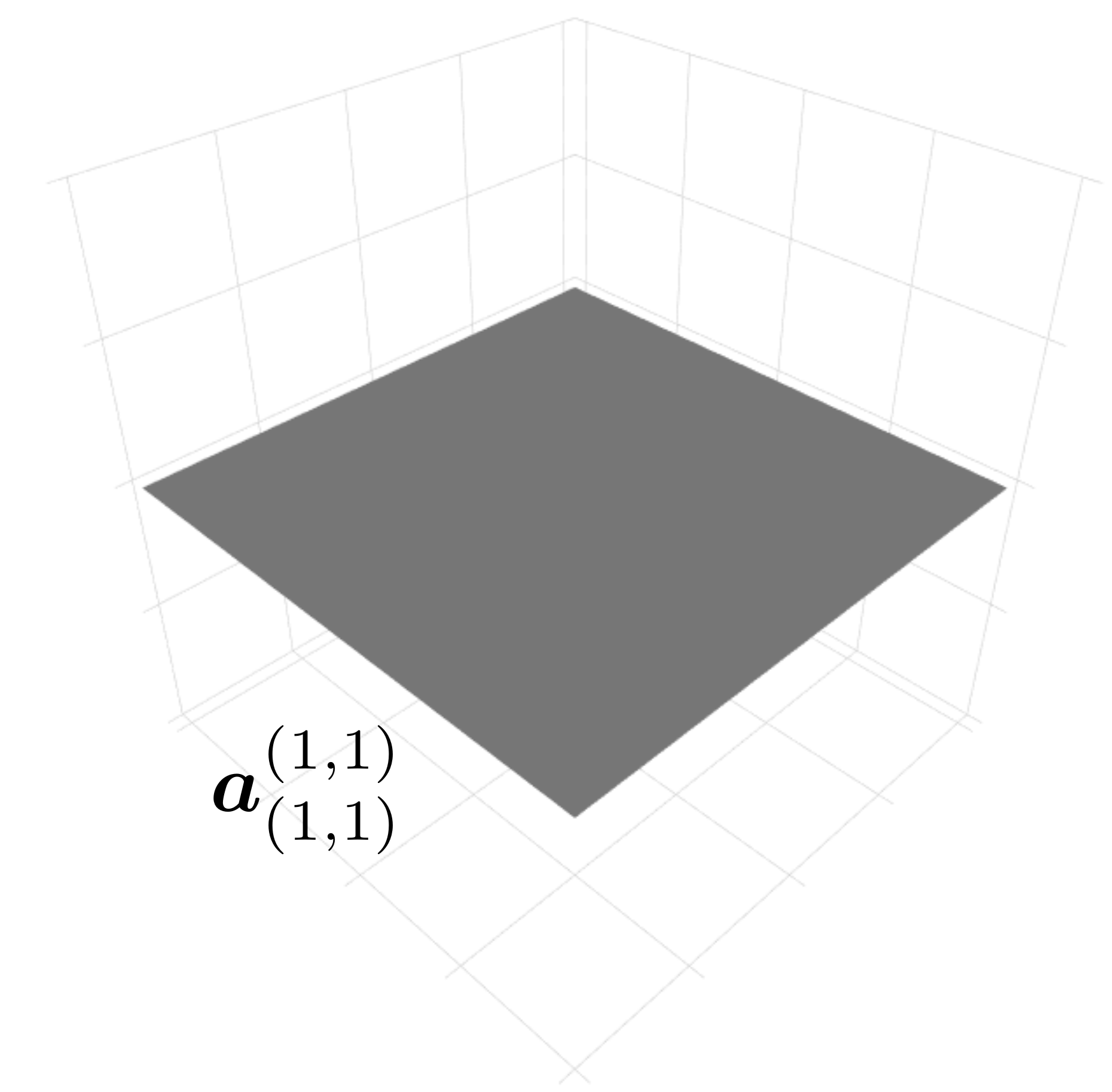} \\
		\hline
		2 & \vspace*{.2cm}\begin{tabular}{ c  c} \includegraphics[scale=.22]{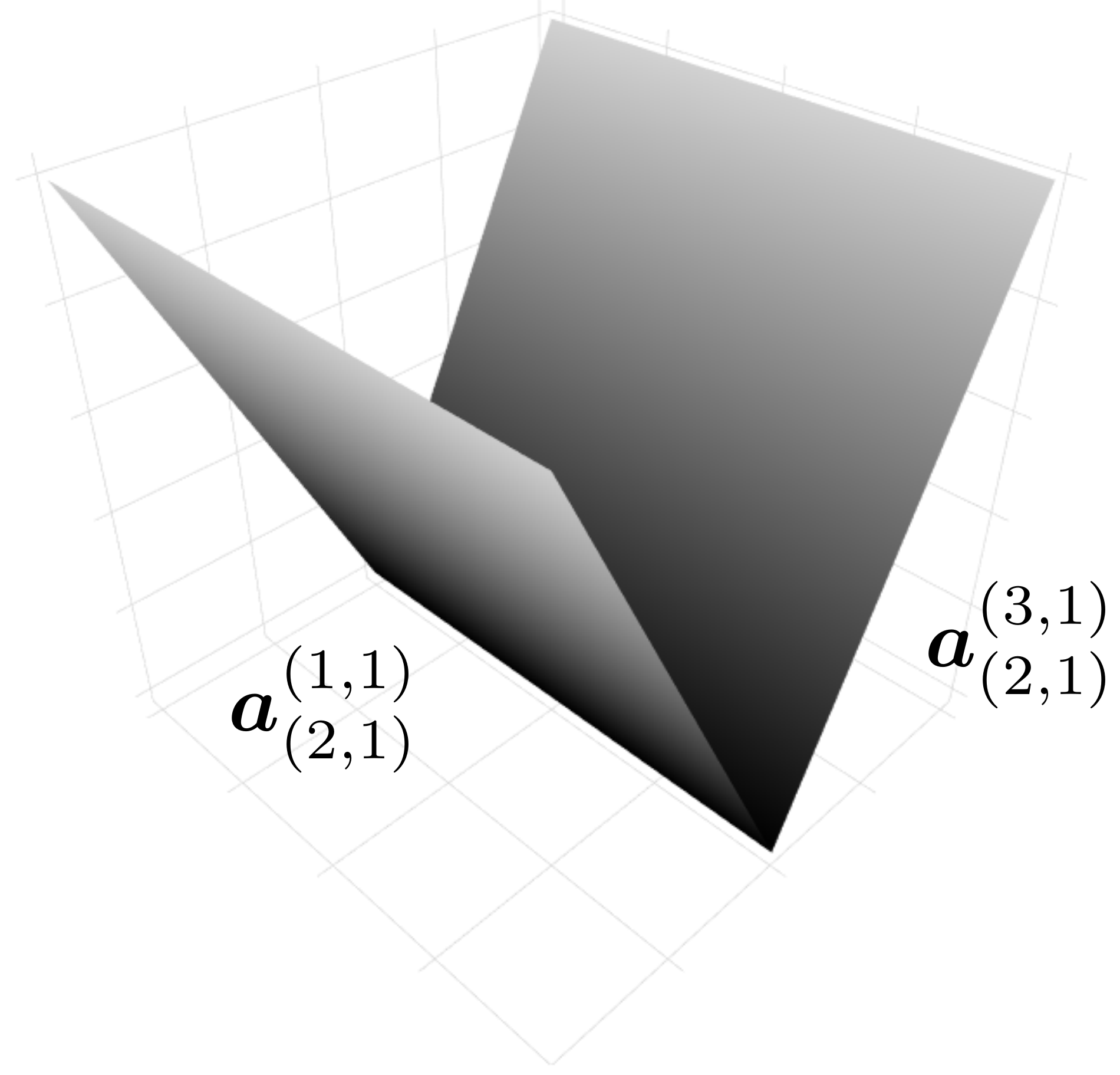} & \includegraphics[scale=.22]{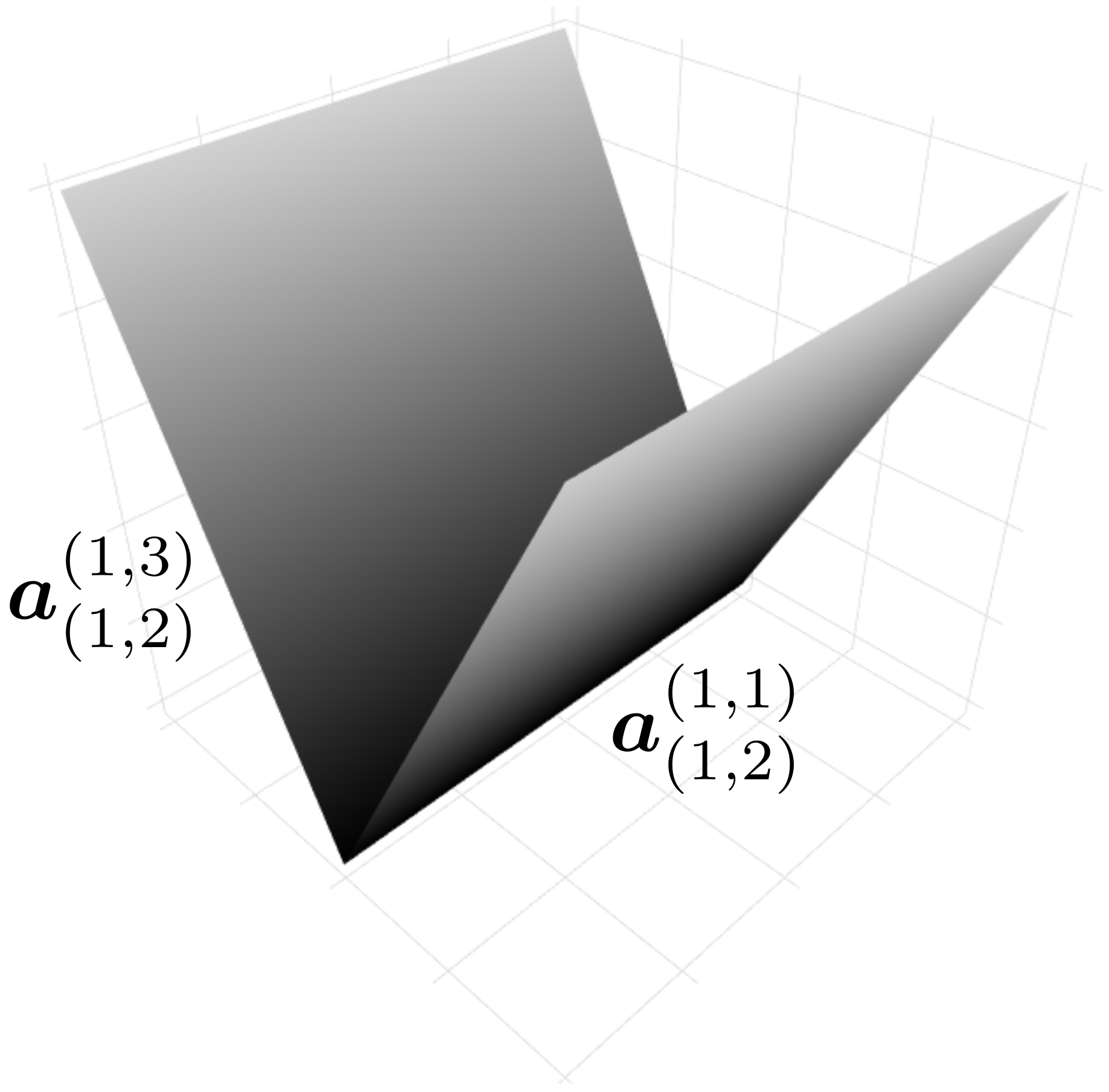} \end{tabular} \\
		\hline
		3 & \vspace*{.2cm}\begin{tabular}{ c  c c} \includegraphics[scale=.22]{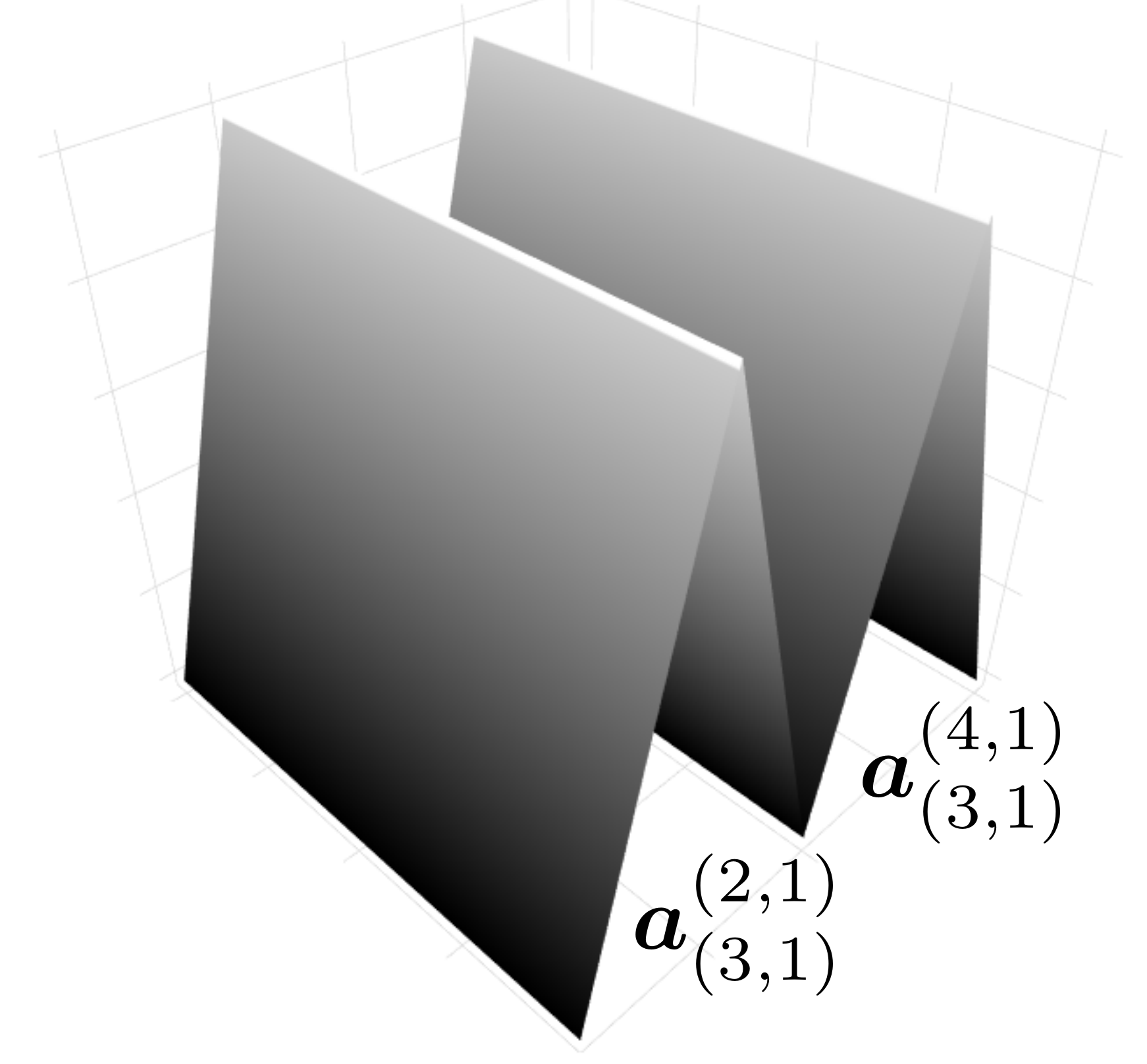} & \includegraphics[scale=.22]{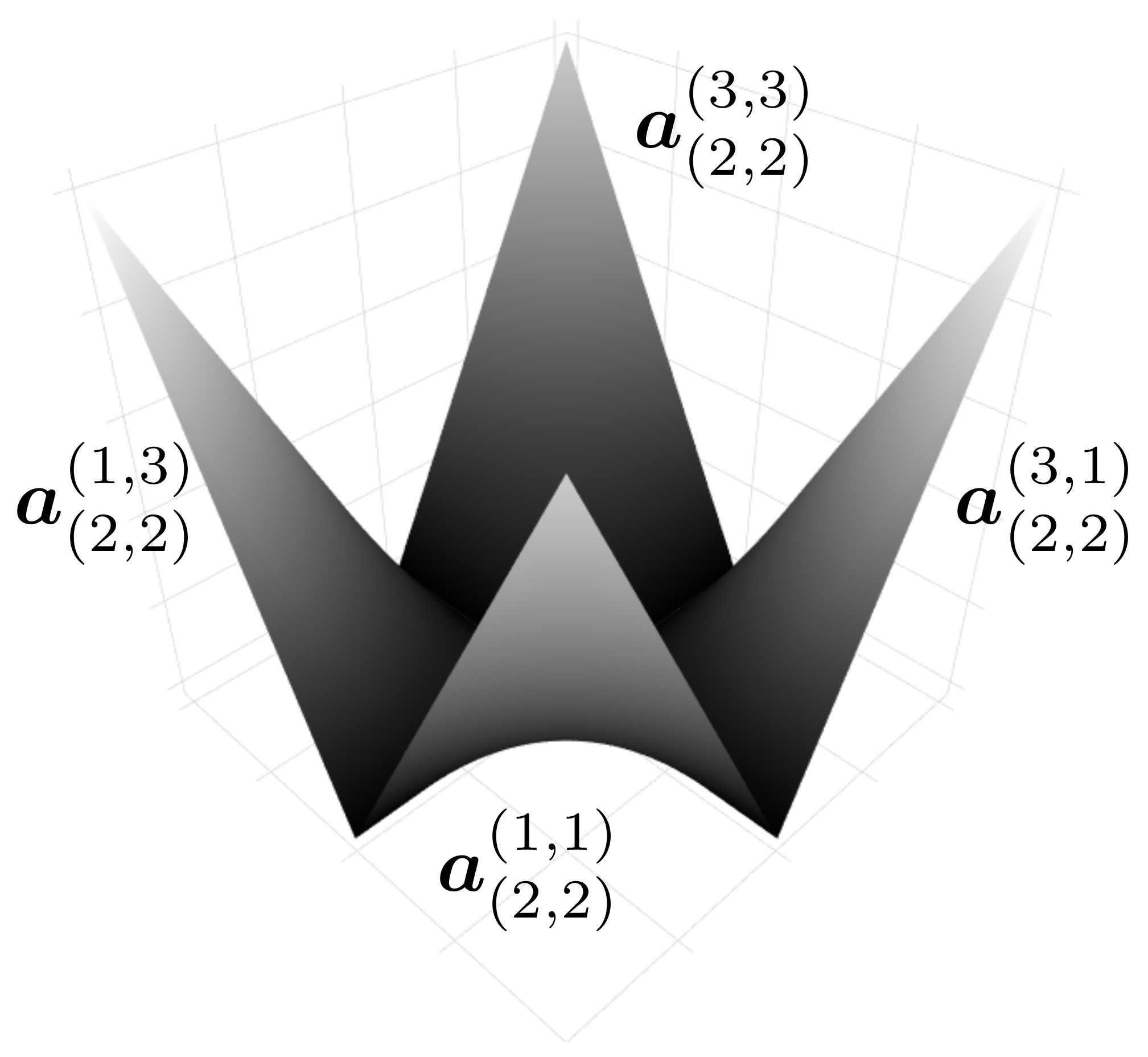} & \includegraphics[scale=.22]{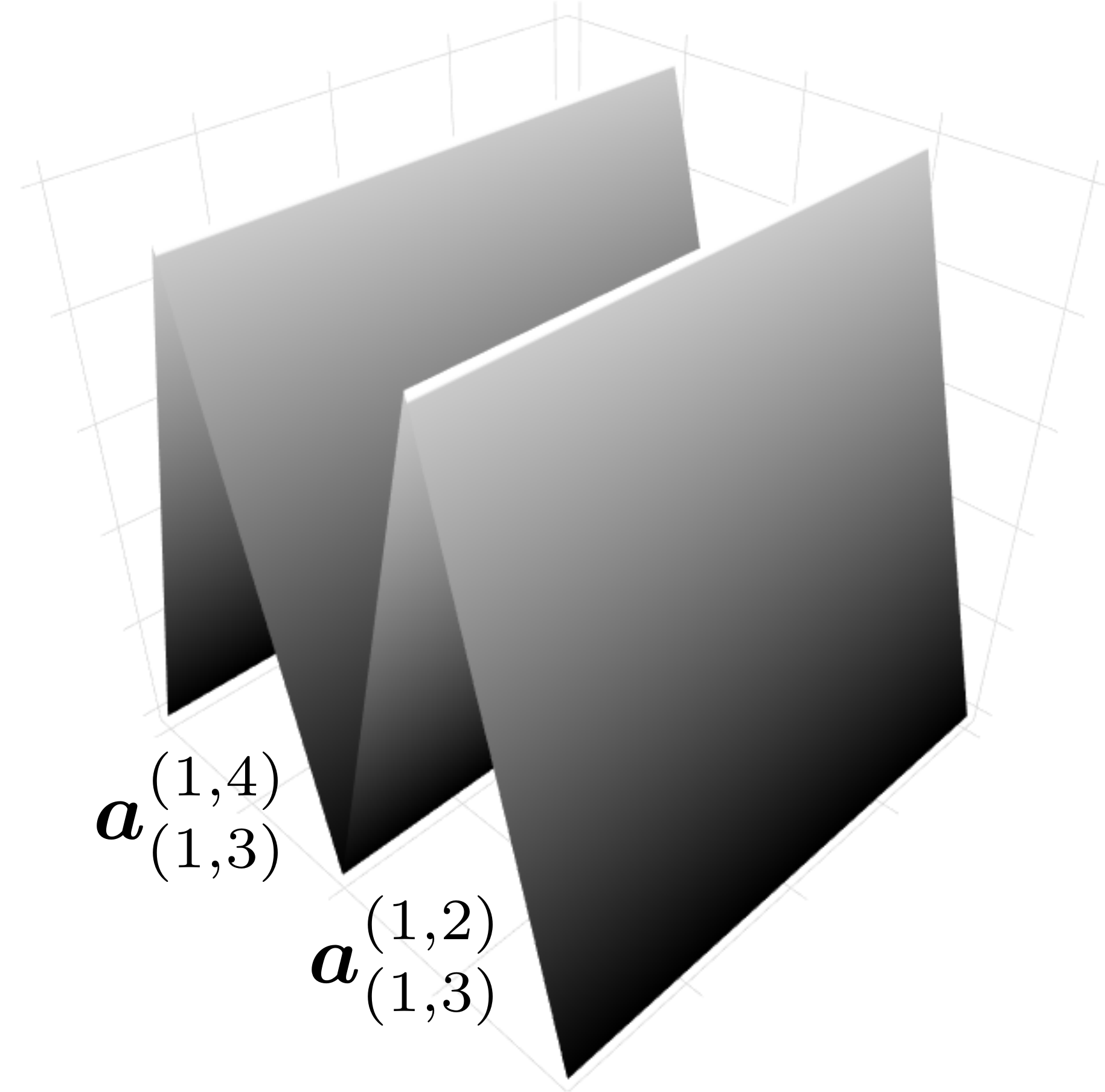} \end{tabular} 
	\end{tabular}
	\vspace{.2cm}
	\caption{Example: Construction of the basis functions $\bs{a}^{\bs{m}}_{\bs{i}}$ in 2D}
	\label{tab:basis}
\end{table}

	\section{Bibliography}
	\bibliography{GM}{}
	\bibliographystyle{unsrtdin1}
	
	\newpage
	
%
%
%
%
%
	

\end{document}